\documentclass[fleqn,usenatbib]{mnras}

\usepackage{newtxtext,newtxmath}

\usepackage[T1]{fontenc}
\usepackage{ae,aecompl}

\usepackage{graphicx}	
\usepackage{amsmath}	
\usepackage{amssymb}	

\usepackage[dvipsnames]{xcolor}

\title[BH Mass of SMSS\,J2157-3602]{A Thirty-Four Billion Solar Mass Black Hole in SMSS\,J2157-3602, the Most Luminous Known Quasar}

\author[Onken et al.]{
Christopher A.\ Onken,$^{1,2}$\thanks{E-mail: \href{mailto:christopher.onken@anu.edu.au}{christopher.onken@anu.edu.au}}
Fuyan Bian,$^{3}$
Xiaohui Fan,$^{4}$
Feige Wang,$^{4,5}$
\newauthor
Christian Wolf,$^{1,2}$
and
Jinyi Yang$^{4}$
\\
$^{1}$Research School of Astronomy and Astrophysics, Australian National University, Canberra, ACT 2611, Australia\\
$^{2}$Centre for Gravitational Astrophysics, Research Schools of Physics, and Astronomy and Astrophysics, Australian National University\\
$^{3}$European Southern Observatory, Alonso de C\'{o}rdova 3107, Casilla 19001, Vitacura, Santiago 19, Chile\\
$^{4}$Steward Observatory, University of Arizona, 933 North Cherry Avenue, Tucson, AZ 85721, USA\\
$^{5}$NASA Hubble Fellow
}

\date{Accepted XXX. Received YYY; in original form ZZZ}

\pubyear{2020}

\begin{document}
\label{firstpage}
\pagerange{\pageref{firstpage}--\pageref{lastpage}}
\maketitle

\begin{abstract}
From near-infrared spectroscopic measurements of the \ion{Mg}{ii} emission line doublet, we estimate the black hole (BH) mass of the quasar, SMSS\,J215728.21\nobreakdash-360215.1, as being \mbox{$(3.4\pm0.6)\times10^{10}$~M$_{\odot}$} and refine the redshift of the quasar to be $z=4.692$. SMSS\,J2157 is the most luminous known quasar, with a 3000\,\AA\ luminosity of (4.7$\pm$0.5)$\times$10$^{47}$~erg\,s$^{-1}$ and an estimated bolometric luminosity of 1.6$\times$10$^{48}$~erg\,s$^{-1}$, yet its Eddington ratio is only $\sim0.4$. Thus, the high luminosity of this quasar is a consequence of its extremely large BH --- one of the most massive BHs at $z>4$.
\end{abstract}


\section{Introduction}

High-luminosity quasars offer important windows into the densest concentrations of baryons at high redshift, as they are likely to reside in well-developed galaxies. The expanding catalogue of such quasars is providing a more complete understanding of black hole (BH) growth in the young Universe \cite[e.g.,][]{2019arXiv191105791I}. The search for the rarest, most luminous quasars and highest-mass BHs continues, facilitated by the huge cosmic volume probed by large-area sky surveys. 

The Sloan Digital Sky Survey \cite[SDSS;][]{2000AJ....120.1579Y} was the first data set in which a large sample of quasars at redshift $\sim 5$ and larger were found \citep{2001AJ....122.2833F,2003AJ....125.1649F,2018A+A...613A..51P}. More recently, the 3pi Steradian Survey of the Panoramic Survey Telescope and Rapid Response System \cite[Pan-STARRS;][]{2016arXiv161205560C} has covered more sky and extended into the Southern hemisphere, which the SkyMapper Southern Survey \cite[SMSS;][]{Onken18} now covers entirely. 

Near-infrared surveys like the UKIRT Infrared Deep Sky Survey \cite[UKIDSS;][]{2007MNRAS.379.1599L}, the UKIRT Hemisphere Survey \cite[UHS;][]{2018MNRAS.473.5113D}, and the VISTA Hemisphere Survey \cite[VHS;][]{VHS} are pushing the frontier of discovery earlier in time, to epochs where the quasar light is nearly all redshifted into the infrared. 

Finally, the Wide-field Infrared Survey Explorer \cite[\textit{WISE};][]{Wright10} and the \textit{Gaia} satellite \citep{2018A+A...616A...1G} play enormously important roles in differentiating the abundant red, foreground stars from genuine high-redshift quasars.

The photometric and astrometric information from SMSS, \textit{WISE}, and \textit{Gaia} has enabled the discovery of 
the most UV-luminous object currently known: the quasar SMSS\,J215728.21\nobreakdash-360215.1 \cite[hereafter, SMSS\,J2157;][]{2018PASA...35...24W,2020MNRAS.491.1970W}.

Based on detailed follow-up, we now report on the BH mass and Eddington ratio of SMSS\,J2157. Section~2 describes the spectroscopic data and measurements of the emission line and continuum properties. In Section~3, we derive the BH mass and Eddington ratio. Section~4 describes imaging data and their constraints on possible gravitational lensing of SMSS\,J2157. In Section~5, we discuss the implications of our measurements.
Throughout the paper, we adopt a flat cosmology with $H_0=70$~km\,s$^{-1}$\,Mpc$^{-1}$ and $\Omega_M=0.3$. 
Photometry is in the AB system unless otherwise indicated.

\section{Spectroscopic Data And Analysis}

With the aim of estimating the BH mass in SMSS\,J2157, we measured the \ion{Mg}{ii} emission line properties as well as the continuum luminosity. Our spectroscopic analysis combined data from two medium-resolution, wide-band spectrographs: Keck/NIRES and VLT/X-shooter. 

\subsection{Keck/NIRES Spectroscopy}

First, we obtained a set of spectra with the Near-Infrared Echellette Spectrometer (NIRES) instrument \citep{2004SPIE.5492.1295W} at the 10-m Keck~2 telescope on UT 2018~June~04. NIRES provides near-complete spectral coverage from 0.94 to 2.45~$\mu$m, with mean resolving power of R$\sim$2700. The sky was clear, and seeing was 0.8~arcsec.
Four 200-s exposures were obtained of SMSS\,J2157, with an ABBA offset pattern.

The data were reduced using PypeIt\footnote{\url{https://github.com/pypeit/PypeIt}}, a Python-based spectroscopic data reduction pipeline \citep{2019zndo...3506873P,2020arXiv200506505P}. The flat fielding and the slit tracing were performed using the flat, following the standard technique. The sky background was subtracted in the A-B or B-A frames using the \citet{2003PASP..115..688K} recipe, in which the sky is fit by an optimal $b$-spline function. Then, the one-dimensional spectra were extracted from the sky-subtracted two-dimensional frames using optimal weighting by inverse variance. The spectral response as a function of wavelength was corrected with spectra of the white dwarf, GD~153, which had been obtained previously. Finally, the spectra were combined and corrected for telluric absorption by fitting the stacked quasar spectrum to the telluric model grids from the Line-By-Line Radiative Transfer Model\footnote{\url{http://rtweb.aer.com/lblrtm.html}} \citep{Clough_etal_2005}. The overall flux normalisation was based on the $J$-band magnitude of SMSS\,J2157 from VHS Data Release 6: 15.65$\pm$0.01 (Vega).

\subsection{VLT/X-shooter Spectroscopy}

We also obtained a set of spectra with the X-shooter instrument \citep{2011A+A...536A.105V} on the European Southern Observatory (ESO) 8-m UT2 Kueyen unit of the Very Large Telescope in Service Mode on UT 2019~October~15 (ESO program 0104.A-0410(A); PI: C.~Wolf). The X-shooter data provide spectral coverage from 3000~\AA\ to nearly 2.5~$\mu$m, with resolving powers of 3200, 5000, and 4300 in the UVB, VIS, and NIR arms of the spectrograph. The two optical spectra used exposure times of 1383 and 1410~s in the UVB and VIS arms, respectively, and were obtained with no binning. The six exposures of the NIR arm used a nodding pattern of 5~arcsec along the slit, with a jitter box size of 0.6~arcsec, and each integration was 480~s long. The data were taken at airmasses between 1.02 and 1.05 in clear but windy weather, and the seeing at 500~nm was estimated to be 1.2~arcsec. The slit widths were 1.6, 1.5, and 1.2~arcsec in the UVB, VIS, and NIR arms, respectively. The spectral response was calibrated by a recent observation of the white dwarf, LTT~7987, obtained as part of the standard X-shooter calibration set regularly updated by ESO. The VIS and NIR arms of the X-shooter spectra were also reduced with PypeIt using the same procedure as for the Keck/NIRES spectra. The UVB spectra were reduced with the standard EsoReflex workflow \citep{2013A+A...559A..96F}.\footnote{\url{http://www.eso.org/sci/software/esoreflex/}}

\subsection{\ion{Mg}{\sc ii} Emission Line}

\begin{figure*}
 \includegraphics[width=17cm]{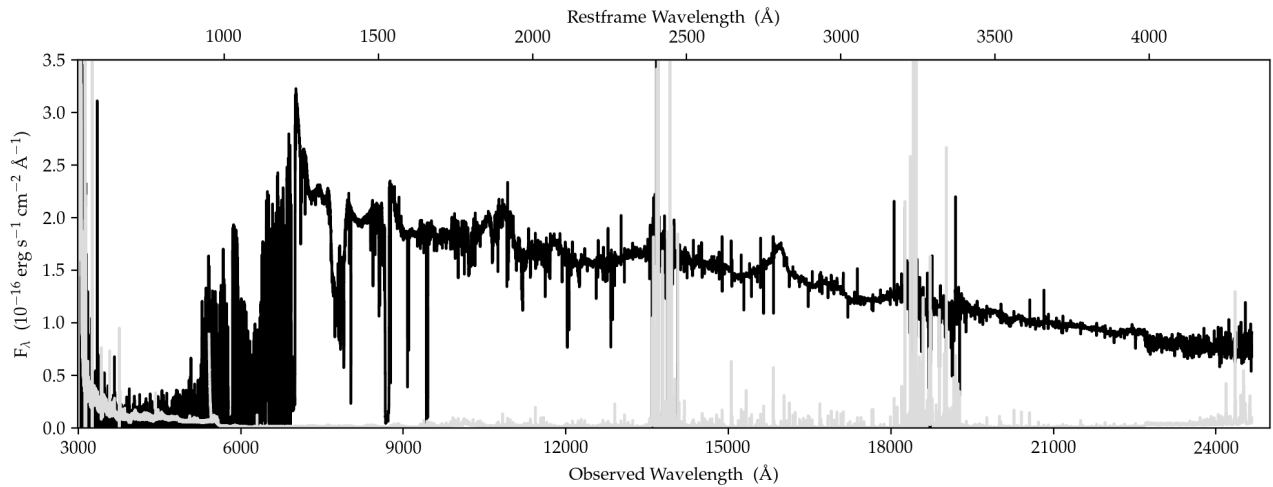}
 \caption{Co-added spectrum of SMSS\,J2157 from Keck/NIRES and VLT/X-shooter. The black line shows the 
 spectrum while the light grey line indicates the error spectrum. Blueward of 1~$\mu$m, only X-shooter contributes to the spectrum. The top axis indicates the restframe wavelength.}
 \label{fig:spec}
\end{figure*}

In order to combine the NIRES and X-shooter spectra, we scaled the NIRES spectrum to the X-shooter flux level using the median in the overlapping wavelength region. We then computed the stacked spectrum in the overlap region after placing the X-shooter spectrum onto the NIRES wavelength grid of 50~km\,s$^{-1}$ per pixel.
The combined X-shooter + NIRES spectrum is shown in Figure~\ref{fig:spec}.

Using the method described by \citet{2011ApJS..194...45S}, the spectrum was fit with a multi-component model that included
a power-law continuum, 
an iron emission-line template \cite[from][]{2001ApJS..134....1V}, and
a \ion{Mg}{ii} profile consisting of a single narrow Gaussian emission line and two broad Gaussian emission lines. The iron template was broadened with a Gaussian of full width at half maximum (FWHM) = 5000~km\,s$^{-1}$. The power-law continuum blueward of \ion{Mg}{ii} is well fit as $(\lambda/\lambda_0)^\alpha$ with $\alpha \approx -0.65$.

\begin{figure}
 \includegraphics[width=\columnwidth]{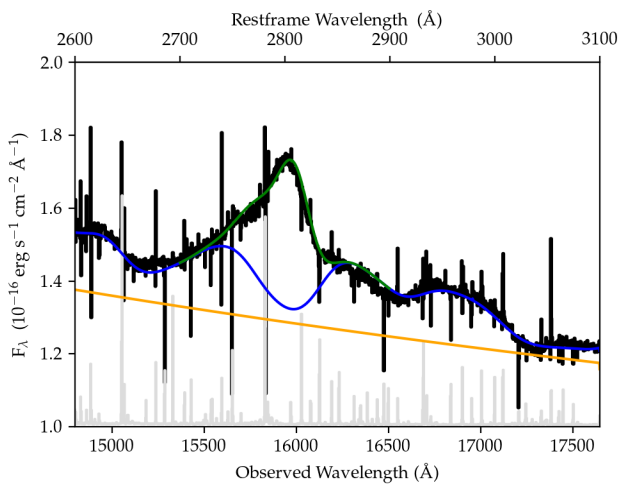}
 \caption{Spectral fit in the region around the \ion{Mg}{ii} emission line, showing the co-added, rebinned spectrum (black), the power-law (PL) fit (orange), the PL + \ion{Fe}{ii} template fit (blue), and the PL + \ion{Fe}{ii} template + \ion{Mg}{ii} emission line fit (green). The error spectrum is plotted in grey, with an additive offset of 1$\times$10$^{-16}$~erg\,s$^{-1}$\,cm$^{-2}$\,\AA. The top axis indicates the restframe wavelength. The \ion{Mg}{ii} FWHM is $5720\pm570$~km\,s$^{-1}$.}
 \label{fig:fit}
\end{figure}

The broad \ion{Mg}{ii} line, modelled by the pair of Gaussians, was found to have a FWHM of $5720\pm570$~km\,s$^{-1}$, where the error is dominated by the uncertainties in the amplitudes of the spectral components underneath the emission line. The spectrum near \ion{Mg}{ii} and the best-fit model are shown in Figure~\ref{fig:fit}. Broad absorption line (BAL) components are visible within the \ion{C}{iv} and Ly~$\alpha$ line profiles\footnote{The BAL features revealed in greater detail by the new, high-S/N spectra make it difficult to assess whether SMSS\,J2157 meets the criteria of \citet{2009ApJ...699..782D} for being a weak-lined quasar.}, but \ion{Mg}{ii} is not significantly affected by such features.

The \ion{Mg}{ii} emission line in bright quasars has been found to reliably trace the systemic redshift, with a small mean offset of $-57$~km\,s$^{-1}$ and a scatter of 205~km\,s$^{-1}$ \citep{2016ApJ...831....7S}. Thus, we refine the systemic redshift of SMSS\,J2157 to be 4.692, within the bounds estimated by \citet{2018PASA...35...24W}. The revised redshift corresponds to a Universe age of 1.247~Gyr, some 20~Myr older than from the original redshift estimate of $z=4.75$.

\subsection{Continuum luminosity}

From the photometrically calibrated near-IR spectroscopy, we are able to obtain a luminosity estimate for SMSS\,J2157 that is free from emission line contributions. 

We find that $\lambda L_{\lambda}$(3000\AA) = $(4.7\pm0.5)\times10^{47}$~erg\,s$^{-1}$, where the uncertainty is estimated by allowing for a reasonable range of \ion{Fe}{ii} emission. This luminosity is consistent with the value in the discovery paper, where the effects of removing the emission line contribution and basing the overall flux calibration on the more precise VHS $J$-band photometry nearly cancel. The revised absolute AB magnitude at 300~nm is $M_{300,\rm AB}=-30.1$. \citet{2018PASA...35...24W} identified PMN\,J1451\nobreakdash-1512 at $z=4.76$ as the next most luminous quasar, and we find that it remains $\approx0.5$~mag fainter than SMSS\,J2157 in $M_{300,\rm AB}$. Additional $z\sim4$ quasars of similar luminosity have recently been discovered,\footnote{We note that LAMOST\,J144757.18+231839.8, catalogued by \citet{2019ApJS..240....6Y}, has been removed from the LAMOST Quasar Survey Catalog following a reclassification of its spectrum as being of "unknown" type (X.-B. Wu, priv. comm.).} i.e., PS1\,J111054.69\nobreakdash-301129.9 \citep{2019ApJ...871..199Y} and PS1\,J212540.96\nobreakdash-171951.4 \citep{2019ApJS..243....5S}. However, based on their near-IR photometry, we find that neither of these quasars are more luminous at restframe 3000~\AA\ than SMSS J2157, which remains the most UV-luminous object known at present.

We can also estimate bluer continuum luminosities from the spectrum. We find values of $\lambda L_{\lambda}$(1350\AA) = $(3.5\pm0.4)\times10^{47}$~erg\,s$^{-1}$ and $\lambda L_{\lambda}$(1450\AA) = $(3.6\pm0.4)\times10^{47}$~erg\,s$^{-1}$. The discovery paper derived $\lambda L_{\lambda}$(1450\AA) = $4.5\times10^{47}$~erg\,s$^{-1}$ (here, revised after applying the new distance correction) from $z$-band VST ATLAS photometry \citep{2015MNRAS.451.4238S}. As the spectroscopic estimates make the assumption that slit losses and flux calibration are consistent across the whole wavelength range, which may not be the case, and the photometric estimate includes contribution from the emission lines (especially  \ion{C}{iv}), we expect that the best estimate of $\lambda L_{\lambda}$(1450\AA) lies somewhere between these values.

\section{BH Mass and Eddington Ratio}

When using \ion{Mg}{ii} to estimate BH masses, one must ensure that the line properties are amenable to this purpose. Due to a dearth of high-quality \ion{Mg}{ii} reverberation mapping results, such BH mass estimates rely on correlations with the large sample of BH masses from H$\beta$ reverberation. 
\citet{2012MNRAS.427.3081T}
show that \ion{Mg}{ii} FWHM values smaller than 6000~km\,s$^{-1}$ accurately trace the H$\beta$ FWHM. This lends a degree of 
confidence
to our ability to estimate the BH mass from the \ion{Mg}{ii} FWHM of 5720~km\,s$^{-1}$ for SMSS\,J2157.

Furthermore, \citet{2008ApJ...689L..13O} characterised a bias in \ion{Mg}{ii} FWHM estimates as a function of Eddington ratio, and presented an empirical correction that represents the FWHM better and traces the BH mass more robustly. For SMSS\,J2157, the \ion{Mg}{ii} FWHM falls in the most reliable part of the FWHM distribution, with no mean bias and a systematic FWHM error of 0.13~dex. We updated this check with the FWHM data from SDSS DR7Q \citep{2011ApJS..194...45S}, which shows that the 1\,800 quasars having \ion{Mg}{ii} FWHMs within the errorbars of our measured value have a median broad-H$\beta$ FWHM of 6280~km\,s$^{-1}$, a difference of less than 1$\sigma$, and that only 10\% have broad-H$\beta$ FWHMs so low as to shift the resulting BH mass down by a factor of two.

As the most luminous quasar currently known, all single-epoch BH mass estimates for SMSS\,J2157 involve extrapolations of the underlying radius---luminosity relationship, and thus are highly sensitive to the adopted slope of that relationship. Because we use the same methodology as \citet{2011ApJS..194...45S} for the line width measurement, we utilise their BH mass equation for \ion{Mg}{ii}, which is based on a correlation between H$\beta$ and \ion{Mg}{ii} BH mass estimates that aims to both anchor itself to the local reverberation mapping determinations from H$\beta$ and be accurate for the high-luminosity quasar sample studied by SDSS. The adopted relation, which has an RMS of 0.25~dex in \citet{2011ApJS..194...45S}, is:
\begin{equation}
  M_{\mathrm{BH}} = 5.5 \left(\frac{L3000}{10^{44} \,\mathrm{erg\,s^{-1}}}\right)^{0.62} \left(\frac{V}{\mathrm{km\,s^{-1}}}\right)^{2}
  \mathrm{M_{\odot}}
\end{equation}
where $V$ is the \ion{Mg}{ii} FWHM, and $L3000$ is the $\lambda L_{\lambda}$ luminosity at 3000~\AA. While noting that the systematic error is likely to be 0.4~dex, we find a BH mass of 
$(3.4\pm0.6)\times10^{10}$~M$_{\odot}$. 

Because of the BAL component, we cannot compare a \ion{C}{iv}-based BH mass estimate for SMSS\,J2157 to the value obtained from \ion{Mg}{ii}\footnote{Other emission lines have sometimes been used for BH mass estimates, including \ion{C}{iii}] $\lambda$1909\AA\ \cite[e.g.,][]{2012ApJ...753..125S,2015ApJ...815..128K}. Because the \ion{C}{iii}] line also exhibits absorption within the line profile, and because such BH measures have been less well calibrated, we refrain from making a \ion{C}{iii}] estimate here.}. However, with the low background levels afforded by the \textit{James Webb Space Telescope}, it will be possible to obtain high-S/N spectra at H$\beta$ (redshifted to 2.77~$\mu$m, inaccessible from ground-based facilities) and H$\alpha$ (3.37~$\mu$m) in very short exposures ($\sim$1~min) with two of the medium-resolution gratings.

Estimating the Eddington ratio requires use of a bolometric correction, here applied to the 3000\,\AA\ continuum luminosity. We adopt the empirical X-ray-to-near-IR corrections of \citet{2012MNRAS.427.1800R}:
\begin{equation}
\log (L_{\mathrm{iso}}) = (1.852 \pm 1.275) + (0.975 \pm 0.028) \log (L3000),
\end{equation}
and $L_{\rm bol} = 0.75 \times L_{\rm iso}$, which give $L_{\rm bol} \approx 1.6\times10^{48}$~erg\,s$^{-1}$, and an Eddington ratio of $\approx$0.4.

\section{Imaging Data and Analysis}

We also obtained new imaging data to place constraints on any magnification of the quasar image by gravitational lensing, as this would bias both the BH mass and the Eddington ratio estimates, increasing them by different factors. Here, we use $J$-band images of SMSS\,J2157 from the slit-viewing camera of Magellan/FIRE, and we analyse publicly available optical images from the Blanco Imaging of the Southern Sky (BLISS) Survey.

\subsection{Magellan/FIRE Near-IR Imaging}

A series of six 30-s $J$-band images were obtained with the slit-viewing camera of the Folded-port InfraRed Echellette (FIRE) instrument \citep{2013PASP..125..270S} at Magellan's 6.5-m Walter Baade Telescope. The dithered images have a plate scale of 0.147~arcsec\,pixel$^{-1}$, and were shifted and co-added, using the centroid of the quasar as the positional reference. Because of the highly variable illumination of the slit viewing camera, it is difficult to flat-field the image, resulting in a large background gradient. However, this does not affect our ability to detect possible extended or multiple images of SMSS\,J2157.

For comparison, we also retrieved archival $J$-band imaging from VHS\footnote{Images and catalogues obtained from the VISTA Science Archive: \url{http://horus.roe.ac.uk/vsa/}}. While the depth of the coadded FIRE image is similar to the VHS image of SMSS\,J2157, the plate scale of FIRE is half that of the VIRCAM instrument used by VHS, and the seeing of 0.6~arcsec is better than the seeing of 0.95~arcsec in the VHS image. 

The FIRE field-of-view is approximately $1\times 1$~arcmin, and from the wider VHS image, only one additional source is expected to appear in the combined FIRE image. That faint source ($J$=20.36~mag per VHS\footnote{We convert the VHS photometry from Vega magnitudes to the AB system by adding 0.91~mag \citep{2007AJ....133..734B}.}) has a signal-to-noise ratio of $S/N \approx 8$, indicating a 5$\sigma$ detection threshold of $\sim 21.0$~mag.

\subsection{Archival Optical Imaging Data}

In addition to the photometric data listed in Table~1 of \citet{2018PASA...35...24W}, we retrieved deep optical imaging data from the BLISS Survey \cite[see details in][]{2019ApJ...875..154M} via the Image Cutout service of the NOAO Data Lab\footnote{\url{https://datalab.noao.edu/}}. BLISS uses the Victor M.\ Blanco 4-m telescope and the Dark Energy Camera (DECam) instrument \citep{2015AJ....150..150F}, which has a plate scale of 0.27~arcsec\,pixel$^{-1}$, and has acquired data in the SDSS $griz$ filters. The deepest of the exposures available for SMSS\,J2157 is a 90-s $i$-band image from UT 2017~July~02. The estimated seeing was 1.0~arcsec, and the 5$\sigma$ detection limit was about 23.3~mag. Here, we use the sky-subtracted, spatially resampled image from the NOAO Data Lab.

\subsection{Constraints on Gravitational Lensing}

Strong gravitational lensing typically operates on angular scales of the order of 1~arcsec and 1~arcmin for galaxies and galaxy clusters, respectively, and could magnify SMSS\,J2157, thus causing an overestimate of its luminosity and BH mass. Here, we explore constraints on that scenario.

\begin{figure}
 \includegraphics[width=\columnwidth]{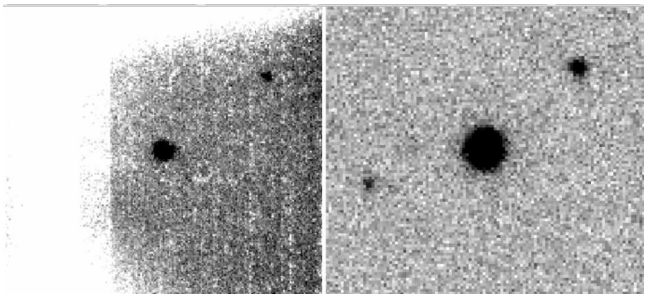}
 \caption{30$\times$30~arcsec image cutouts around SMSS\,J2157 in the $J$-band (\textit{left}, from Magellan/FIRE) and the $i$-band (\textit{right}, from BLISS). In both images, SMSS\,J2157 has a point-source spatial profile, with no indications for multiple images or a lensing galaxy.}
 \label{fig:images}
\end{figure}

The galaxy luminosity function at $z=1-1.5$, where the gravitational lensing likelihood peaks for sources at $z=4-5$, has a restframe $V$-band (observed-frame $J$-band) $M^{*}_{V}$ value of $-22.2$~mag, corresponding to $J\approx 21.5$~mag \citep{2012ApJ...748..126M}. Thus, 
our $J$-band imaging is expected to reach about 0.5~mag brighter than the $M^{*}_{V}$ value for the galaxy redshifts most likely to bias our luminosity estimate. Within the FIRE $J$-band image (Figure~\ref{fig:images}, left), the only two sources detected are those expected from the existing VHS dataset (SMSS\,J2157 and the fainter source, 12~arcsec away). With only one additional source in the field-of-view, distortions from a point source profile for SMSS\,J2157 cannot be tightly constrained from the FIRE image, but there is no indication of multiple peaks in the spatial distribution. Thus, no candidate lenses are identified in the $J$-band data.

For sources at $z=1.5$, the BLISS $i$-band image probes 3000\,\AA\ restframe emission, and we can obtain an estimate of the sensitivity to potential lensing galaxies from the 1500\,\AA\ luminosity function \cite[e.g.,][]{2010ApJ...725L.150O}. With an $M^{*}_{1500}$ around $-20$~mag, the BLISS images probe to about 1~mag above $M^{*}_{1500}$, here with a particular sensitivity to star-forming galaxies at that redshift.
The BLISS image (Figure~\ref{fig:images}, right) shows the two sources from the $J$-band image (SMSS\,J2157 and the fainter source at $i$=21.3~mag), and one additional source, 11~arcsec from SMSS\,J2157. The new source is also faint, with $i$=22.3~mag. As with the $J$-band image, there are no indications of the presence of lensing galaxies, and the spatial profile of SMSS\,J2157 exhibits no signatures of being anything other than a point source.

Using the BLISS $i$-band image, we can constrain the presence of an unresolved second image of SMSS\,J2157 arising from strong gravitational lensing. We adopt a simple singular isothermal sphere lens model at $z=1.5$, which makes predictions for the magnification ratio between the two images, as well as the total magnification, for a given angular offset between lens and source (in units of the Einstein radius). Thus, for a given magnification ratio, we aim to determine the maximum offset between the two images of SMSS\,J2157 that could go undetected, which we take as being twice the Einstein radius. In detail, we shift the image by an integer number of pixels in the East-West direction, coadd the shifted image to the original with a given value of the magnification ratio, and run {\sc Source Extractor} \citep{1996A+AS..117..393B} on the resulting image. The shape parameters for the doubly imaged sources are then compared to the distribution of object shapes from the original image, including SMSS\,J2157.

For image magnification ratios up to 5, which imply a total magnification of the source of a factor of 3 or more, we find that the Einstein radius must be no larger than 1~pixel (0.27~arcsec), or else the real shape parameters of SMSS\,J2157 become a clear outlier in the flux--FWHM plane. This implies a lens mass of $\lesssim 4\times10^{10}$~M$_{\odot}$, likely having an associated stellar mass less than $\sim10^{8}$~M$_{\odot}$ \cite[e.g.,][]{2020A+A...634A.135G}. Photometric detection of such a small galaxy, similar in scale to the Large Magellanic Cloud \citep{2018ApJ...864...55Z}, would be extremely challenging at $z=1.5$, particularly with a coincident luminous quasar. At higher magnification ratios, where the second image is {\it de}-magnified, the limits on the Einstein radius are progressively weaker, but the effect of the lensing is also less substantial.

Within the SMSS imaging data, we can look for wavelength-dependent position shifts as another indicator of a foreground lensing galaxy. What we find is that, amongst the filters having detections of SMSS\,J2157 ($g, r, i, z$), the average positions vary by less than the typical systematic RMS shift between the SMSS coordinates and those from {\it Gaia} DR2 (0.16~arcsec), again providing no evidence of a lensing galaxy.

From the spectroscopic data, one might also hope to see evidence of absorption from any interstellar or circumgalactic gas in the vicinity of a foreground lens. The most common such absorption is from \ion{Mg}{ii}, but at a redshift of $z\lesssim1.5$, the associated spectral features would appear blueward of $\sim$7000~\AA, which is already so heavily affected by absorption associated with SMSS\,J2157 that identifying foreground lines is intractable. 

Finally, gravitational lensing would enhance the Eddington ratio by approximately the square root of the magnification. However, as noted above, the measured Eddington ratio of $\approx$0.4 is not significantly larger than that observed for other luminous quasars.

For any given quasar, one cannot exclude the possibility of a precisely aligned low mass lens, but as the sample of high-redshift quasars grows, a better understanding of the luminosity function and the BH mass function will help to reveal significant outliers. In conclusion, none of the imaging datasets available raise a concern about strong gravitational lensing of SMSS\,J2157, and we currently have no reason to believe that the luminosity (and consequently, both the BH mass and the Eddington ratio) are over-estimated.

\section{Discussion}

The measured Eddington ratio of 0.4 places SMSS\,J2157 within the typical distribution of luminous quasars \cite[e.g.,][]{2006ApJ...648..128K,2017ApJS..228....9K}, implying that the extreme nature of its luminosity is a consequence of its extraordinarily high BH mass.

In order to make fair comparisons with the other $z>4$ quasars having BH mass estimates of $\gtrsim10^{10}$~M$_{\odot}$, we first recalculate their masses with the same methodology and scaling relations as above. For J0306+1853 at $z=5.363$, the \ion{Mg}{ii} and $L3000$ data of \citet{2015ApJ...807L...9W} give a BH mass estimate of $2.0\times10^{10}$~M$_{\odot}$, $L_{\rm bol}=7.2\times10^{47}$~erg\,s$^{-1}$, and an Eddington ratio of 0.3.
For J0100+2802 at $z=6.3$, the measurements of \citet{2015Natur.518..512W} give a BH mass of $2.1\times10^{10}$, $L_{\rm bol}=1.1\times10^{48}$~erg\,s$^{-1}$, and an Eddington ratio of 0.4. As with SMSS\,J2157, any systematic errors significantly outweigh the statistical uncertainties for the BH mass and Eddington ratio, as evidenced by the factor of $\sim2$ change in those values simply by adopting different scaling relations\footnote{We omit a small number of high-mass, high-redshift BHs from this discussion because of smaller masses indicated by \ion{Mg}{ii} \citep{2015MNRAS.450L..34G, 2014ApJ...795L..29Y}, conflicting mass estimates from emission lines other than \ion{Mg}{ii} \citep{2015ApJ...799..189Z}, or a lack of \ion{Mg}{ii}-based estimates \citep{2007AJ....134.1150J, 2015ApJ...806..109J, 2017ApJS..231...16J, 2019ApJ...873...35S}. Given the primacy of the systematic errors, adding a further layer of systematics to enable such mass comparisons is unlikely to be informative.}. 

What these measurements suggest is that between $z=6.3$ and $z=4.692$, from a Universe age of 0.86~Gyr to 1.25~Gyr, the accretion properties of the most massive BHs do not change substantially. As that time span is of order 10 Salpeter ({\it e}-folding) times, and the mass difference is only $\sim50$\%, it is the highest-redshift BH of this set that places the strongest constraint on the seed mass and early growth rate of these immense BHs. In fact, the less-massive BHs discovered at $z>7$ \citep{2011Natur.474..616M,2018Natur.553..473B} place stronger limits still, and continue to challenge models \citep{2019MNRAS.485.2694A}. However, moving beyond individual, extreme objects, we believe that fundamental progress in understanding the early growth of supermassive BHs will ultimately come from obtaining large and complete samples.

\section*{Acknowledgements}

We thank the referee for their comments, all helpful. We thank Indrajit Patra for prompting a more thorough check of other luminous quasars, and Xue-Bing Wu for looking into the LAMOST Quasar Survey data. We acknowledge support from the Australian Research Council through Discovery Project DP190100252. XF and JY acknowledges supports from NASA ADAP grant NNX17AF28G S003 and US NSF grant AST 19-08284. FW acknowledges support by NASA through the NASA Hubble Fellowship grant \#HST-HF2-51448.001-A awarded by the Space Telescope Science Institute, which is operated by the Association of Universities for Research in Astronomy, Incorporated, under NASA contract NAS5-26555. The national facility capability for SkyMapper has been funded through ARC LIEF grant LE130100104 from the Australian Research Council, awarded to the University of Sydney, the Australian National University, Swinburne University of Technology, the University of Queensland, the University of Western Australia, the University of Melbourne, Curtin University of Technology, Monash University and the Australian Astronomical Observatory. 

We thank the support staff at Keck, Magellan, and the VLT. This paper includes data gathered with the 6.5 meter Magellan Telescopes located at Las Campanas Observatory, Chile. This paper is based, in part, on observations made with ESO Telescopes at the La Silla Paranal Observatory under programme ID 0104.A-0410(A). Some of the data presented herein were obtained at the W.M. Keck Observatory, which is operated as a scientific partnership among the California Institute of Technology, the University of California and the National Aeronautics and Space Administration. The Observatory was made possible by the generous financial support of the W.M. Keck Foundation. The authors wish to recognise and acknowledge the very significant cultural role and reverence that the summit of Mauna Kea has always had within the indigenous Hawaiian community. We are most fortunate to have the opportunity to conduct observations from this mountain. 

This project used data obtained with the Dark Energy Camera (DECam), which was constructed by the Dark Energy Survey (DES) collaboration. Funding for the DES Projects has been provided by the US Department of Energy, the US National Science Foundation, the Ministry of Science and Education of Spain, the Science and Technology Facilities Council of the United Kingdom, the Higher Education Funding Council for England, the National Center for Supercomputing Applications at the University of Illinois at Urbana-Champaign, the Kavli Institute for Cosmological Physics at the University of Chicago, Center for Cosmology and Astro-Particle Physics at the Ohio State University, the Mitchell Institute for Fundamental Physics and Astronomy at Texas A\&M University, Financiadora de Estudos e Projetos, Funda\c{c}\~{a}o Carlos Chagas Filho de Amparo \`{a} Pesquisa do Estado do Rio de Janeiro, Conselho Nacional de Desenvolvimento Cient\'{i}fico e Tecnol\/{o}gico and the Minist\/{e}rio da Ci\^{e}ncia, Tecnologia e Inova\c{c}\~{a}o, the Deutsche Forschungsgemeinschaft and the Collaborating Institutions in the Dark Energy Survey. The Collaborating Institutions are Argonne National Laboratory, the University of California at Santa Cruz, the University of Cambridge, Centro de Investigaciones En\'{e}rgeticas, Medioambientales y Tecnol\'{o}gicas-Madrid, the University of Chicago, University College London, the DES-Brazil Consortium, the University of Edinburgh, the Eidgen\"{o}ssische Technische Hochschule (ETH) Z\"{u}rich, Fermi National Accelerator Laboratory, the University of Illinois at Urbana-Champaign, the Institut de Ci\`{e}ncies de l'Espai (IEEC/CSIC), the Institut de F\'{i}sica d'Altes Energies, Lawrence Berkeley National Laboratory, the Ludwig-Maximilians Universit\"{a}t M\"{u}nchen and the associated Excellence Cluster Universe, the University of Michigan, the NSF's National Optical-Infrared Astronomy Research Laboratory, the University of Nottingham, the Ohio State University, the OzDES Membership Consortium, the University of Pennsylvania, the University of Portsmouth, SLAC National Accelerator Laboratory, Stanford University, the University of Sussex, and Texas A\&M University. Based on observations at Cerro Tololo Inter-American Observatory, NSF's National Optical-Infrared Astronomy Research Laboratory (NSF's OIR Lab Prop. 2017A-0260; PI: M.\ Soares-Santos), which is managed by the Association of Universities for Research in Astronomy (AURA) under a cooperative agreement with the National Science Foundation. This research uses services or data provided by the Astro Data Lab at NSF's National Optical-Infrared Astronomy Research Laboratory. NSF's OIR Lab is operated by the Association of Universities for Research in Astronomy (AURA), Inc. under a cooperative agreement with the National Science Foundation. This paper uses data from the VISTA Hemisphere Survey ESO programme ID: 179.A-2010 (PI. McMahon). The VISTA Data Flow System pipeline processing and science archive are described in \citet{2004SPIE.5493..411I}, \citet{2008MNRAS.384..637H}, and \citet{2012A+A...548A.119C}. We acknowledge use of the Ned Wright's web-based Cosmology Calculator \citep{2006PASP..118.1711W}. This work made use of Jupyter notebooks \citep{Kluyver:2016aa}.

\bsp	
\label{lastpage}
\end{document}